\documentclass[fleqn,twoside]{article}
\usepackage{latexsym,amsmath}
\begin{document}
\newcommand{\bbox}{\stackrel{-}{\Box}}
\newcommand{\bnabla}{\bar {\nabla}}
\begin{center}
INFLATIONARY DILATONIC de SITTER UNIVERSE   FROM     SUPER
YANG - MILLS  THEORY PERTURBED BY SCALARS AND SPINORS\\
\bigskip
Iver Brevik\footnote{E-mail:  iver.h.brevik@mtf.ntnu.no}\\
Department of Energy and Process Engineering\\
Norwegian University of Science and Technology\\
N-7491 Trondheim, Norway\\

\bigskip

John Quiroga\footnote{E-mail: jquiroga@tspu.edu.ru}\\
Department of Physics\\
Universidad Tecnol\'ogica de Pereira\\
Colombia\\
and\\
Lab. for Fundamental Study,\\ Tomsk State Pedagogical
University, \\Tomsk
634041, Russia\\
\end{center}
\bigskip
\begin{center}
August 2003\\
\end{center}
\begin{abstract}
In this paper a quantum $\cal{N}$  = 4  super Yang-Mills theory
perturbed by dilaton-coupled scalars and spinor fields, is
considered. The induced effective action for such a theory is
calculated on a dilaton-gravitational background using the
conformal anomaly found via the  AdS/CFT correspondence. Considering
such an effective action (using the large N method) as a quantum
correction to the classical gravity action with cosmological
constant, we study the effect from the dilaton on the scale factor
(this corresponds to an inflationary universe without dilaton ).
It is shown that, depending on the initial conditions for the
dilaton, the dilaton may slow down, or accelerate, the inflation
process. At late times, the dilaton is decaying exponentially.
Different possible cases corresponding to a dilatonic dS Universe are
analyzed with respect to the equations of motion.
\end{abstract}
\newpage

Recent evidence has shown that the present Universe is
subject to accelerated expansion  and may thus be entering an
inflationary phase now. In view of this, taking into account
that the inflationary Universe is currently considered to be a 
realistic model for the evolution of the very Universe 
(for a general review, see
\cite{KT}), one may think about 
reconsidering quantum cosmology and  constructing a new (or
modified) version of the theory of the very early Universe.

In the present work, we consider one theory of this kind which has
become very popular recently  in connection with the AdS/CFT
correspondence, namely  quantum cosmology as following from $\cal
{N}$ = 4 quantum super YM theory.  Using conformal anomaly on a
dilaton-gravitational background the anomaly induced effective
action was constructed in \cite{brevik99}, and the consequences it
may lead to in the early Universe were discussed in
the same reference. The present paper presents the generalization
of such a model for quantum cosmology \cite{brevik99} where ${\cal N}\;\;$ =4 
SYM is perturbated by quantum scalar and spinor fields interacting
with the dilaton. Taking into account that on a purely gravitational
background such an effective action leads to the possibility of
inflation, we will show that the role of the dilaton is to
accelerate, or to slow down, the inflationary expansion, depending
on which choice is made for the initial conditions of the dilaton.

 Let us start from the Lagrangian of local superconformally
 invariant  $\cal {N}$ = 4  super YM  theory in the background of
 $\cal{N}$ = 4  conformal supergravity. The corresponding vector multiplet
is
 $(A_\mu, \psi_i, X_{ij})$. Supposing that super  YM  theory
 interacts with conformal supergravity in a  SU(1,1)  covariant
 way and keeping only kinetic terms, we get:
\begin{eqnarray}
L_{SYM}&=&-\frac{1}{4}(e^{-\phi}F_{\mu\nu}F^{\mu\nu}+\tilde{C}F^{\mu\nu} {F^*}_{\mu
\nu})- \nonumber  \\
       & &-\frac{1}{2}\bar{\psi}^i \gamma^\mu D_\mu \psi^i-
\frac{1}{4}X_{ij}(-D^2+\frac{1}{6}R)X^{ij}+...
\label{1}\end{eqnarray}

We note that the scalar $\phi$ from the conformal
supergravity multiplet is written as $\tilde{C}+ie^{-\phi}$. We
also  note that the first term in (\ref{1}) describes the
dilaton coupled electromagnetic field whose conformal anomaly has
been found in \cite{NO}.  $\phi$ is a complex scalar (the dilaton).

As we intend to include scalars and spinors in our theory, we must 
add appropriate terms in the action (1). Assuming that there are $M$  $4d$ dilaton coupled
 scalars  $X^k$, we write

\begin{equation}
L=f(Re\,\phi)g^{\mu\nu}\partial_{\mu}X^k\partial_{\nu}X^k \,, \qquad
k=1,...,M. \label{scalar}\end{equation}

Here, $f$ is taken to be some function of the real real part of the dilaton.

Now, as matter Lagrangian we take the one associated with $M$
massless (Dirac) spinors, i.e.
\begin{equation} \label{matterOA}
L_s=e^{A\phi}\sum_{i=1}^M \bar\psi_i\gamma^\mu\nabla_\mu\psi^i\ .
\end{equation}

It is interesting to note that from the above action one can construct dilaton-
coupled
Wess-Zumino theory \cite{SJG}.
 On a purely bosonic background with only non-zero gravitational and dilaton
fields, the conformal anomaly for ${\cal N}$ = 4 super YM theory
has been calculated in \cite{NO1} via AdS/CFT correspondence
\cite{MA} to be (adding conformal anomaly for dilaton coupled
scalar);
\begin{eqnarray}
T&=&b(F+\frac{2}{3}\Box R)+b'G+b''\Box R +C[\Box \phi^*\Box
\phi-2(R^{\mu\nu}-\frac{1}{3}g^{\mu\nu}
R)\nabla_{\mu}\phi^*\nabla_{\nu}\phi)]+\nonumber
\\&+&a_1\frac{[(\nabla f)(\nabla f)]^2}{f^4}+a_2 \Box
\left(\frac{(\nabla f)(\nabla f)}{f^2}\right).
\label{2}\end{eqnarray}

Here
\[
b=\frac{N^2-1}{(4\pi)^2}\frac{N_s+6N_f+12N_v}{120}+\frac{M}{120(4\pi)^2}=
\frac{N^2-1}{4(4\pi)^2}+\frac{M}{120(4\pi)^2}, \]
\[b'=-\frac{N^2-1}{(4\pi)^2}\frac{N_s+11N_f+62N_v}{360}-\frac{M}{360(4\pi)^2
}=-\frac{N^2-1}{4(4\pi)^2}-\frac{M}{360(4\pi)^2}, \]
\[ C=\frac{N^2-1}{(4\pi)^2}N_v=\frac{N^2-1}{(4\pi)^2}. \]
\[a_1=\frac{M}{32(4\pi)^2}\;\;,\qquad a_2=\frac{M}{24(4\pi)^2}\]

In the above expression for the anomaly we have taken into account
the fact that  $N_s=6$, $N_f=2$, $N_v=1$ in ${\cal N}$ = 4 SU(N)
super YM theory; \(
F=R_{\mu\nu\alpha\beta}R^{\mu\nu\alpha\beta}-2R_{\mu\nu}R^{\mu\nu}+\frac{1}{
3}R^2$
is the square of the Weyl tensor in four dimensions; $G$ is the
Gauss-Bonnet invariant. The prefactor $N^2-1$ appears because all
fields are in the adjoint representation. The conformal anomaly
for four-dimensional dilaton-coupled scalar has been found in
refs. \cite{SNSD}.

It is important to note that the anomaly (\ref{2}) was obtained
adding the anomaly for dilaton coupled scalar, but that the terms for
the dilaton coupled spinor field were not considered since such
an expression  would be too complicated to write down explicitly.

 Note that both $N$ and $M$ may be considered to be big
parameters. So, one can study large-$N$ or large-$M$ expansions.

Let us now find the anomaly induced effective action \cite{RFT}
(for a review, see \cite{BOS}). We will write it in non-covariant,
local form:
\begin{eqnarray}
W   &  =    &  b \int d^4 x \sqrt{-\bar{g}} \bar{F} \sigma+
   b'\int d^4 x \sqrt{-\bar{g}} [ \, \sigma [2\,{\bbox}^2 +\nonumber \\
    &  +    &     4 \bar{R}^{\mu \nu} {\bnabla}_\mu {\bnabla}_\nu
   -\frac{4}{3}\bar{R} \bbox  +\frac{2}{3}({\bnabla}^\mu
\bar{R}){\bnabla}_\mu ]\sigma
     +(\bar{G}- \frac{2}{3} {\bbox} \bar{R})\sigma ]  -\nonumber \\
    & -  & \frac{1}{12}[b''+\frac{2}{3}(b+b')] \int d^4 x \sqrt{-\bar{g}}[
\,
\bar{R} -6 \bbox \sigma -6( {\bnabla}_\mu \sigma )({\bnabla}^\mu \sigma) ]^2
+\nonumber \\
    &  + & C \int d^4 x \sqrt{-\bar{g}}\, \sigma \phi^* ( {\bbox}^2 +2
{\bar{R}}^{\mu \nu}
{\bnabla}_\mu {\bnabla}_\nu-\frac{2}{3} \bar{R} \bbox
+\frac{1}{3}({\bnabla}^\mu \bar{R} ){\bnabla}_\mu)\phi + \nonumber
\\ &  + & \int d^4 x \sqrt{-\bar{g}}\, \{ a_1\frac{[(\nabla
 f)(\nabla f)]^2}{f^4}\sigma + a_2 \Box
\left(\frac{(\nabla f)(\nabla f)}{f^2}\right)\sigma +\nonumber
\\ & + &a_2
\frac{(\nabla f)(\nabla
f)}{f^2}\left[(\nabla\sigma)(\nabla\sigma)\right]\}.
\label{4}\end{eqnarray}

The computation of the anomaly--induced EA for the dilaton coupled
spinor field has been done in \cite{NNO}, and the result, in the
non-covariant local form, reads:

\begin{eqnarray} \label{vii} \lefteqn{\hspace{-.8cm} W_s=\int d^4x
\sqrt{-\bar g} \bigg\{\tilde{b} \bar F \sigma_1 + 2\tilde{b'}
\sigma_1\Big[ 
{\bbox}^2
 + 2 \bar
R^{\mu\nu}\bar\nabla_\mu\bar\nabla_\nu
- \frac{2}{3}\bar R{\bbox} + \frac{1}{3}(\bar\nabla^\mu\bar
R)\bar\nabla_\mu \Big]\sigma_1} \nonumber \\
&&\!\!\!\!\!\!\! 
+\, \tilde{b'}\sigma_1\Big(\bar G -\frac{2}{ 3}{\bbox} \bar{R}\Big)
-\frac{1}{18}(\tilde{b} + \tilde{b'})\left[\bar R - 6 {\bbox}
\sigma_1 - 6(\bar\nabla_\mu \sigma_1)(\bar\nabla^\mu \sigma_1)
\right]^2\bigg\}\,, \end{eqnarray}

Here, $\sigma_1=\sigma+ A\phi /3$, for Dirac spinors
$\tilde{b}=\frac{3M}{ 60(4\pi)^2}$, $\tilde{b'}=-\frac{11 M }{ 360
(4\pi)^2}$.

Moreover, $V_3$ is the (infinite) volume of 3-dimensional flat space,
 $'\equiv{d / d\eta}$, and $\sigma=\ln a$ where $a(\eta)$ is the
scale factor.

Note that in the conformal anomaly (\ref{2}) we used $ g_{\mu\nu}
= e^{2 \sigma} \bar{g}_{\mu \nu} $, and all quantities in
(\ref{4}) are calculated with the help of the overbar metric.

Since we know that the anomaly induced effective action is defined
with accuracy up to a conformally invariant functional, we may
limit ourselves to a conformally flat metric, i. e. $\bar{g}_{\mu
\nu}= {\eta}_{\mu \nu} $. In this case, the conformally invariant
functional on a purely gravitational background is zero, and  $
W+W_s $ in Eq. (\ref{4}) gives the complete contribution to the
one-loop effective action.  In addition to this we will assume
that only the real part of the dilaton coupled to SYM theory is
non-zero .

The anomaly induced effective action (\ref{4}) may now be
simplified significantly (due to the fact that $ {\bar{g}}_{\mu
\nu} = {\eta}_{\mu \nu} $):
\begin{eqnarray}
W   &  =    & \int d^4 x \{ \, 2 b' \sigma {\Box}^2 \sigma -3 ( b''
+\frac{2}{3}(b+b')) \times \nonumber \\
    &   & \times (\Box \sigma + \partial_\mu \sigma \partial ^\mu \sigma)^2
+ C \sigma \,\phi \, {\Box}^2 \phi + a_1\frac{[(\nabla
 f)(\nabla f)]^2}{f^4}\sigma  +\nonumber \\ &  & + a_2 \Box
\left(\frac{(\nabla f)(\nabla f)}{f^2}\right)\sigma +a_2
\frac{(\nabla f)(\nabla
f)}{f^2}\left[(\nabla\sigma)(\nabla\sigma)\right]\},
\label{5}\end{eqnarray}

where all derivatives are now flat ones.

Moreover,  the anomaly-induced EA for dilaton coupled
spinor field, eq.~(\ref{vii}), is
 \begin{equation}
\label{1ef} W_s=V_3\int d\eta \left\{2\tilde{b'} \sigma_1
\sigma_1'''' - 2(\tilde{b} + \tilde{b'})\left( \sigma_1'' -
{\sigma_1'}^2 \right)^2\right\}\,.
\end{equation}

Considering the case when the scale factor $ a(\eta) $ depends
only on conformal time: $\sigma(\eta)=\ln a(\eta) $, one has to
add the anomaly induced effective action to the classical
gravitational action:
\begin{equation}
S_{cl} =- \frac{1}{\kappa} \int d^4 x \sqrt{-g}\, (R+6\Lambda) =
-\frac{1}{\kappa}\int d^4 x e^{4\sigma}
(-6e^{-2\sigma}((\sigma')^2+(\sigma''))+6\Lambda),
\label{6}\end{equation}

where $ \kappa = 16 \pi G $.

We now consider the complete action, $S_{total}=S_{cl}+W+W_s$, and
find the gravitational equation of motion by taking the
variational derivative with respect to the scale factor
$a=e^\sigma$. Similarly the field equation of motion is found by
taking the variational derivative with respect to $\phi$.

Now, the equations of motion for the action $S_{total}=S_{cl}+W+ W_s$,
assuming that $\sigma $ and $ \phi $ depend only on the conformal
time $\eta$, may be written in the following form (assuming the
simplest choice $f(Re~\phi)=\phi$):
\begin{eqnarray}\label{eqmov}
\!\!\!\!\!\!\!\!&&24(\tilde b + \tilde
b')\left(\frac{a'}{a}+\frac{A\phi'}{3}\right)^2-2(3b''+2b+2\tilde
b)\frac{a''''}{a} + 8(3b''+2b) \frac{
a'\,a'''}{a^2}+\nonumber \\
          &  & +6(3b''+2b+2\tilde b)\frac{
{a''}^2}{a^2}-2(3b''+2b)\left(6-\frac{12b'}{3b''+2b}+\frac{24\tilde
b}{3b''+2b}\right)\frac{a''\,{a'}^2}{a^3} -\nonumber
\\ & &
- 24(b'-\tilde b) \frac{{a'}^4}{a^4}+ \frac{12}{\kappa } a\,a'' -
\frac{24\Lambda}{\kappa }a^4-4\tilde b\frac{A\phi''''}{3}+
C\phi\,\phi''''+ a_1\frac{{\phi'}^4}{\phi^4}+
a_2\left[\frac{{\phi'}^2}{\phi^2}\right]''-\nonumber\\
& &-
2a_2\left(\frac{a''}{a}-\frac{{a'}^2}{a^2}\right)\frac{{\phi'}^2}{\phi^2}-
4a_2\frac{a'}{a}\frac{\phi'\phi''}{\phi^2}+4a_2\frac{a'}{a}\frac{{\phi'}^3}
{\phi^3}=0,
\nonumber \\
              &   & \nonumber \\
              &   &\frac{A}{3}\left[24(\tilde b + \tilde
              b')\left(\frac{a'}{a}+\frac{A\phi'}{3}\right)^2-4\tilde{
b}\left(\frac{a''''}{a}-4\frac{a'\,a'''}{a^2}-3\frac{{a''}^2}{a^2}+12\frac{{a'}^2
\,a''}{a^3}-6\frac{{a'}^4}{a^4}+\right.\right.
              \nonumber\\ &
&+\left.\left.\frac{A\phi''''}{3}\right)\right]+C\left[ \ln a \; \phi'''' +
(\ln{a}\;
\phi)''''\right]-4a_1\frac{a'}{a}\frac{{\phi'}^3}{\phi^4}+2a_2\frac{a''}{a}
\left(\frac{{\phi'}^2}{\phi^3}-\frac{\phi''}{\phi^2}\right)+\nonumber
              \\ &&+a_2\left(2\frac{a'a''}{a^2}-\frac{a'''}{a}\right)\frac{\phi'}{\phi^2}-
12a_1\left[\frac{{\phi'}^2\phi''}{\phi^4}-\frac{{\phi'}^4}{\phi^5}\right]\ln
              a=0,
\end{eqnarray}
 where $3 b''+2 b \neq 0$. The natural choice for $b''$ is to take $b''=0$,
since the
choice of $b''$ does not make any difference in the physical
effects.

Performing the analysis of these equations one may consider some
interesting cases. Firstly we may consider case when in our theory
we have scalar but the spinor is absent. For this problem, the  solution
was obtained in \cite{quiroga03}. For this case, an approximate
special solution of eqs. (\ref{eqmov}) may be obtained, when the term with $\ln a$
is negligible. We got $a(t)=a_0e^{Ht}$, $\ln a
\sim Ht$. Moreover, $H$ is proportional to the Planck mass, so
$Ht$ is small quantity and such an approximation is justified. So
using the ansatz (for case of no dilaton it corresponds to anomaly driven
inflation \cite{Sta})

\begin{equation}
a(t) \simeq \tilde{a}_0 e^{\tilde{H}t},~~~~\phi(t) \simeq \phi_0
e^{-\alpha \tilde{H}t} \label{11}\end{equation}
special solutions were found. Since we know that both $N$ and $M$
may be considered to be big parameters, it means  one can study
large-$N$ or large-$M$ expansions. So in the performed analysis
we considered two possibilities for the parameters, namely
$N^2\approx M$ and $M\gg N^2$. It was found that in both cases,
the obtained solution was bigger than that obtained in the absence
of the dilaton \cite{brevik} so we conclude that, in this case, the
role of the dilaton is to make the inflation faster, as compared
with the case of no dilaton.

The second interesting case occurs when we
neglect the scalar and take into account only the effect of the
spinor. In such a case, as in the case above, one may search for approximate
  equations of motion (\ref{eqmov})
(see references \cite{NNO,geyer}), by using the same ansatz
(\ref{11}). From the form of these equations one finds that
 these solutions will  depend on the parameters of
 the spinor and the dilaton. So one may conclude that such a quantum
 dS Universe, perturbed by spinors only, may represent the real world if one chooses
  these parameters appropriately. In fact, related problems were solved
 in \cite{NNO,geyer}. In these references, by using an ansatz similar
 to (\ref{11}),  approximate solutions were obtained from
 which  one observes that the value of $H^2$ significantly increases
 by the contribution from the dilaton. One 
  special solution was found explicitly that describes a Brans-Dicke
 non-singular Universe with a (much slower) expanding dilaton.
 It is important to note that this is a purely quantum solution
 which does not exist at the classical level.

 Returning to the general theory, i.e. the situation when
 both scalar and spinor are present, eqs. (\ref{eqmov}) fully describe
 our quantum FRW Universe. These equations are too complicated to be solved 
 analytically so it becomes necessary to use some
approximation or
 numerical methods. As in the above cases one may search for
 special approximate solutions by using the same ansatz
 (\ref{11}). As in the two cases before, we find numerical solutions
 describing a dilatonic dS Universe. Clearly, the parameters of such a 
Universe are defined by the parameters for the scalar, the spinor, and
 of course  the dilaton. The important lesson of this study is that by a fine
tuning of the values of the theory
 parameters one may find  that quantum dS
 Universe occurs. Such a Universe has attracted interest  recently in connection
with the dS/CFT correspondence (see \cite{sdo} for a recent review), related to the  
current acceleration of the Universe.

Finally, let us make a comment on the form of the coupling function $f$. We assumed above the simplest possible form, $f(Re~\phi)=\phi$. Another simple possibility, assuming $\phi$ to depend only on conformal time, $\phi=\phi(\eta)$, would be to assume an exponential form:
\begin{equation}
f(\eta)=e^{-\beta \eta},
\end{equation}
with $\beta$ a constant. With this expression for $f$, the anomaly induced effective action (7) becomes 
\[
 W=\int d^4 x\,\Big\{ 2b' \sigma \sigma''''-3\left(b''+\frac{2}{3}(b+b')\right) 
(\sigma''+{\sigma'}^2)^2 \]
\begin{equation}
+C\sigma \phi \phi'''' +a_1 \,\beta^4 \sigma +a_2\,\beta^2 {\sigma'}^2 \Big\}.
\end{equation}
This form is still analytically tractable. As above, the gravitational equation of motion is found by taking the variational derivative with respect to $a=e^\sigma$, whereas the field equation  is found by taking the variational derivative with respect to $\phi$.

\end{document}